\documentclass[11pt,a4paper,twoside]{article}
\usepackage{amssymb,amscd}
\usepackage{times,cite,epsfig}
\newcommand{\be}{\begin{eqnarray}}
\newcommand{\ee}{\end{eqnarray}}
\newcommand{\bez}{\begin{eqnarray*}}
\newcommand{\eez}{\end{eqnarray*}}
\newcommand{\pa}{\partial}
\newcommand{\A}{{\mathcal A}}
\renewcommand{\H}{{\mathcal H}}
\renewcommand{\d}{{\rm d}}
\newcommand{\Dirac}{\slash \!\!\!\! D}
\newcommand{\eqref}{\ref}

\begin{document}

\title{Noncommutative Geometries and Gravity\thanks{Talk presented 
at the Third Mexican Meeting on Mathematical and Experimental Physics, 
Symposium on Gravitation and Cosmology, Mexico City, 10-14 September 2007. 
To appear in the proceedings.} }

\author{Folkert M\"uller-Hoissen \\
  Max-Planck-Institute for Dynamics and Self-Organization, 
  37073 G\"ottingen, Germany}

\date{}

\maketitle

\begin{abstract}
We briefly review ideas about ``noncommutativity of space-time'' and 
approaches toward a corresponding theory of gravity.
\end{abstract}

\noindent
\textbf{PACS:} 02.40.Gh, 04.50.+h, 04.60.-m \\
\textbf{Keywords:} Noncommutative geometry, space-time, deformation, gravity

\section{Introduction}
``Noncommutative geometry'' (NCG) is a broad framework in which notions 
of space, symmetry and (differential) geometry can be generalized in 
various ways. In this short review we will concentrate on aspects 
related to the concepts of space-time and gravity. 
Let us recall that spaces can be traded for commutative rings or algebras.
Relativists and geometers are familiar with this point of view: in a 
chart on a manifold (e.g. space-time) one works with the commutative 
algebra of functions generated by coordinates 
$x^\mu$.\footnote{An algebraic formulation of General Relativity 
has already been proposed in 1972 by R. Geroch \cite{Gero72,Land+Marm91}.}
\footnote{More technically, given a locally compact 
space $M$, the set of continuous $\mathbb{C}$-valued 
functions on it (that vanish ``at infinity'' if $M$ is not compact) 
becomes a commutative $C^\ast$ algebra with the $L^\infty$-norm and 
$f^\ast$ the complex conjugate of $f$.  
Furthermore, every commutative $C^\ast$-algebra $\A$ is isomorphic to the  
algebra $C(M)$ of continuous functions on some locally compact space $M$ 
(Gelfand-Naimark theorem, see e.g. \cite{Land97,GVF01}). This 
involves the construction of a space (``Gelfand spectrum'') 
as the set of non-zero characters, i.e. homomorphisms into $\mathbb{C}$. 
In case of the algebra of continuous functions on a Hausdorff space, 
one recovers the original space. $M$ is compact if $\A$ is unital. }

If a commutative associative algebra thus corresponds to a topological space, 
a \emph{non}commutative associative\footnote{Even \emph{nonassociative}  
algebras are of interest. In particular nonassociative star products 
appear in string theory \cite{Corn+Schi02}. But we will leave this aside. } 
algebra $\A$ may be regarded as a ``noncommutative space''.\footnote{
$C^\ast$-algebras are particularly nice since they admit a faithful 
representation by bounded operators on a Hilbert space. 
In quantum physics, a familiar example of a noncommutative $C^\ast$-algebra 
is the (Weyl) algebra of one-parameter unitary groups generated by position 
and momentum operators, but more flexible are ``resolvent algebras''  \cite{Buch+Grun07}. }  
The analogue of a vector bundle (needed to formulate gauge theory) on such 
a noncommutative space is then a module over $\A$.\footnote{This is based 
on the equivalence of vector bundles over a compact space $M$ and finitely 
generated projective modules over $C(M)$ (Serre-Swan theorem). 
See \cite{GVF01}, for instance.} 

A rule which associates with a ``commutative space'' some noncommutative space 
is a kind of quantization, analogous to canonical quantization in 
physics, which replaces an algebra of functions on a phase 
space with a Heisenberg (Weyl) algebra of operators on a Hilbert space, 
or deformation quantization \cite{BFFLS78a,BFFLS78b}, 
which deforms the commutative product of functions to the noncommutative 
Groenewold-Moyal product \cite{Groe46,Moya49}. Further 
examples of ``quantized spaces'' are provided by quantum groups that 
are deformations of classical groups reformulated as Hopf algebras 
(see \cite{Klim+Schm97}, for example). Several noncommutative spaces 
do play a role in physical models and theories. The idea of 
``noncommutative space-time'' is more speculative, however.\footnote{An 
example is Snyder's ``quantized space-time'' which originates from the 
five-dimensional de Sitter space regarded as ``momentum space'' of a 
particle \cite{Snyd47}. It preserves Lorentz invariance, but breaks 
translational invariance (see also \cite{Yang47}). More generally 
curved momentum spaces correspond to noncommutative configuration spaces, 
see \cite{Mats+Well98} for the example of a point particle in 
$(2+1)$-dimensional gravity.} 
Let us discuss critically three arguments that appear in the literature 
in favor of it. Others will be addressed in the following sections. 
\vskip.1cm

\noindent
1. Before renormalization theory had been developed, quantum field theory (QFT)  
was plagued by apparently uncontrollable infinite expressions. 
In those days the idea came up that noncommutativity of coordinates 
could help to eliminate these (ultraviolet) divergences 
\cite{Snyd47}. Meanwhile the believe is that QFT 
on noncommutative spaces (with an \emph{infinite} number of degrees of freedom) 
still requires renormalization \cite{Filk96,Vari+Grac99}. 
But for a non-renormalizable theory like 
perturbative Einstein gravity on Minkowski space, improvements  
(comparable with that of string theory) could perhaps be achieved in such a 
way.\footnote{It should also be noticed that ultraviolet divergences appear 
in {\em integrated} expressions and therefore already the introduction 
of a weaker kind of noncommutativity, namely a noncommutativity 
between (commuting) functions and differentials can do a good job 
\cite{DMHS93b}. }
\vskip.1cm

\noindent
2. At least operationally the concept of 
space-time underlying General Relativity does not make sense below the 
length scale given by the Planck length $\ell_P = \sqrt{\hbar \mathcal{G}/c^3}$ 
(where $\mathcal{G}$ is Newton's gravitational constant).
In order to resolve space (-time) with greater accuracy we need more energy. 
A resolution limit is then obtained when the radius of the ball 
into which the energy is transmitted becomes smaller 
than the corresponding Schwarzschild radius, in which case no information 
can escape from this area (see e.g. \cite{DFR94,DFR95}, and \cite{Gara95} 
for related arguments).\footnote{In string theory a resolution limit is 
given by the string length.} 
This suggests space-time uncertainty relations, 
which can be realized \cite{DFR94,DFR95} by turning coordinate 
functions into noncommuting self-adjoint operators: 
\be
  [ \hat{x}^\mu , \hat{x}^\nu ] = i \, Q^{\mu\nu}  \; .
\ee
In a low energy approximation, the operators $Q^{\mu\nu} = - Q^{\nu\mu}$ 
should be negligible and $\hat{x}^\mu$ become inertial coordinates. 
Assuming covariance under the Poincar{\'e} group, treating 
$Q^{\mu\nu}$ as a tensor, the analysis in \cite{DFR94,DFR95}
led to conditions for $Q^{\mu\nu}$, which are in particular 
satisfied if $Q^{\mu\nu}$ is a central element of the algebra, 
subject to some algebraic constraints. 
A word of caution is in place, however. 
In General Relativity coordinates are not regarded as observables, 
all the information about space-time resides in the metric tensor. 
Space-time uncertainties may then result from quantization of the 
metric (on a commutative space). In contrast, the ``coordinates'' used in 
\cite{DFR94,DFR95} are assumed to carry metric information like 
inertial coordinates in Special Relativity. 
\vskip.1cm

\noindent
3. A kind of space-time noncommutativity appears in string theory in the 
so-called Seiberg-Witten limit \cite{Seib+Witt99}. The bosonic part of the 
(open) string action in a background metric $g_{\mu\nu}$ and background 
$B$-field is 
\be
   S = \frac{1}{4 \pi \alpha'} \int_\Sigma g_{\mu\nu} \, \pa_a X^\mu \, 
         \pa^a X^\nu \, \d^2 \sigma 
      - \frac{i}{2} \int_\Sigma B_{\mu\nu} \, \d X^\mu \wedge \d X^\nu \; .
\ee
If $|g_{\mu\nu}| \ll |\alpha' B_{\mu\nu}|$ with \emph{constant} $B_{\mu\nu}$, 
then $S \approx -(i/2) B_{\mu\nu} \int_{\pa \Sigma} X^\mu \,  \pa_t X^\nu$ 
(with $\pa_t$ tangential to the world sheet boundary $\pa \Sigma$), 
which upon canonical quantization leads to 
\be 
     [\hat{X}^\mu , \hat{X}^\nu ] = i \, \theta^{\mu\nu}  \qquad 
     \mbox{on } \pa\Sigma  \, ,     \label{ncMink}
\ee
where $\theta^{\mu\nu} = (B^{-1})^{\mu\nu}$.\footnote{More precisely, 
here we should consider a space-filling $D$-brane, or a lower-dimensional 
$Dp$-brane, then split the set of coordinates accordingly and assume 
maximal rank of $B$, see \cite{Seib+Witt99}.} 
Thus the embedding functions 
$X^\mu$ restricted to the string end points become noncommuting 
operators in this limit. This heuristic derivation  
very much parallels that of noncommutative coordinates in the case of 
the Landau problem of a quantum particle in a plane with perpendicular 
strong external magnetic field (see \cite{Jack02,Hein+Ilde07} for instance). 
This does not mean that the classical space-time somehow disappears, 
but rather that in certain situations physics is (more) effectively 
described in terms of certain noncommutative ``coordinates''. 
\vskip.1cm

There is an important advantage of noncommutativity (e.g. noncommutative 
space-time) as compared with discretization (``discrete space-time''). 
Whereas discretization, i.e. replacing the continuum by a discrete space, 
typically breaks continuous symmetries, noncommutativity is more 
flexible.\footnote{Introducing noncommutativity can actually restore 
continuous symmetries which got lost by discretization. 
Discretizing the sphere by reducing it to a north and a south pole 
obviously destroys its continuous symmetries. The 
remaining freedom can be expressed by the set of diagonal $2 \times 2$ 
matrices, on which $SO(3)$ can only act trivially. But if we extend 
it to the noncommutative space of \emph{all} $2 \times 2$ matrices, there 
is a non-trivial action of $SO(3)$. See also \cite{Mado99ped} and the 
fuzzy sphere example. } 

\noindent
\footnotesize
\emph{Example} \cite{Mado92}. Let $J_a$, $a=1,2,3$, be a Hermitian basis of $su(2)$ such that 
$[J_a,J_b] = i \, \epsilon_{abc} \, J_c$. In the $j$-dimensional 
irreducible representation, the value of the Casimir operator is 
given by $J_1^2 + J_2^2 + J_3^2 = \frac{j^2-1}{4} \, I$
(with the unit matrix $I$). Then
$x_{(j)a} := 2 \, r \, J_a/\sqrt{j^2-1}$, $a=1,2,3$,  
with a positive real constant $r$, satisfy $\vec{x}_{(j)}^2 = r^2 \, I$, 
which formally corresponds to the equation defining the two-dimensional 
sphere in three-dimensional Euclidean space. Since 
$[x_{(j)a} , x_{(j)b}] = (2ir/\sqrt{j^2-1}) \, \epsilon_{abc} \, x_{(j)c}$, 
the algebra becomes commutative in the limit $j \to \infty$, and indeed approximates 
the sphere. $SU(2)$ acts by conjugation (adjoint representation) on its Lie algebra 
and thus on the \emph{fuzzy sphere} $S^2_j$ (i.e. the algebra generated by $x_{(j)a}$, 
$a=1,2,3$), preserving the ``sphere constraint''.
\normalsize
\vskip.1cm

An algebra alone is not sufficient to describe a space-time, we need 
an additional structure which encodes the metric information. 
There are several (mathematical) ways to implement this, some 
of which will be considered in later sections. 

In the following sections we gather some essentials from 
several approaches toward a ``noncommutative'' generalization 
of the notions of space-time and gravity. It is based on a certain 
(surely personally based) 
selection from the existing literature and we regret for not being 
able to give consideration to all of those who contributed to this field.

\section{Moyal-deformed space-time and gravity}

\paragraph{Moyal deformation of $\mathbb{R}^n$}
In deformation quantization \cite{BFFLS78a,BFFLS78b}, a noncommutative 
algebra is obtained by replacing the commutative product of 
functions by the (Groenewold-) Moyal product \cite{Groe46,Moya49} 
(so that the Poisson bracket is replaced by the Moyal bracket 
\cite{Moya49}), defined for functions on $\mathbb{R}^n$ 
in terms of coordinates $x^\mu$ by 
\be
    f \star h := \mathbf{m}_{\mathcal{F}}(f \otimes h) \, , 
      \quad 
    \mathbf{m}_{\mathcal{F}} := \mathbf{m} \circ \mathcal{F}^{-1} \, , 
      \quad
     \mathcal{F} 
   := \exp\Big( -\frac{i}{2} \theta^{\mu\nu} \pa_\mu \otimes \pa_\nu \Big) \; . 
           \label{Moyal-prod}
\ee
Here $\theta^{\mu\nu}$ are real antisymmetric constants and 
$\mathbf{m}(f \otimes h) = fh$. In particular, we have 
$x^\mu \star x^\nu = x^\mu \, x^\nu + (i/2) \theta^{\mu\nu}$ and thus 
\be
   [ x^\mu , x^\nu ]_\star := x^\mu \star x^\nu - x^\nu \star x^\mu 
                            = i \, \theta^{\mu\nu} \, ,  \label{Moyal-Rn}
\ee
which makes contact with (\ref{ncMink}). Indeed, the Seiberg-Witten 
limit of string theory can be described in terms of the Moyal product. 
Clearly the above relation and also 
the $\star$-product of two scalars are invariant under constant linear 
(e.g. Lorentz) transformations if $\theta^{\mu\nu}$ are treated as 
tensor components. Note that complex conjugation is an involution:
$(f \star h)^\ast = h^\ast \star f^\ast$. 
For Schwartz space functions, 
$\int f \star h \, \d x^n = \int f \, h \, \d x^n$.

\paragraph{Kontsevich star-product}
There is a covariantization of the Moyal-product and moreover a generalization 
to the case where $\theta^{\mu\nu}$ is an arbitrary Poisson tensor field 
\cite{Kont03,Catt+Feld00}. In local coordinates we have 
\be
   f \star h &=& f h + \frac{i}{2} \theta^{\mu\nu} \pa_\mu f \, \pa_\nu h 
     - \frac{1}{8} \theta^{\mu \kappa} \theta^{\nu \lambda} \pa_\mu \pa_\nu f 
      \, \pa_\kappa \pa_\lambda h \nonumber \\
  && - \frac{1}{12} \theta^{\mu \lambda} \pa_\lambda \theta^{\nu \kappa} 
       ( \pa_\mu \pa_\nu f \, \pa_\kappa h - \pa_\nu f \, \pa_\mu \pa_\kappa h) 
     + \mathcal{O}(\theta^3) 
\ee
(see also \cite{Szab06}). 
This is indeed associative due to the Jacobi identity of the Poisson structure 
(which is equivalent to 
$\theta^{\kappa [\lambda} \pa_\kappa \theta^{\mu \nu]}=0$). Kontsevich found a 
formal combinatorial expression to all orders in $\theta$ \cite{Kont03} 
(see also \cite{Catt+Feld00}). 
Under a change of coordinates the $\star$-product changes by an equivalence transformation $f \star' h = \mathcal{S}^{-1}(\mathcal{S}f \star \mathcal{S}h)$ 
with an operator $\mathcal{S}$ (see also \cite{Cham01,CTZZ06,Szab06}). 
In string theory a non-constant Poisson tensor originates from a non-constant 
$B$-field on a $D$-brane. See \cite{Szab06} for corresponding examples.

\paragraph{Twisted Poincar{\'e} symmetry}
Let us think of $x^\mu$ as (inertial) space-time coordinates. 
Instead of regarding the parameters $\theta^{\mu\nu}$ in the Moyal-product 
as tensor components, let us try to treat them as fixed \emph{constant} 
numbers, so we may restrict to space-space noncommutativity 
by setting $\theta^{0 \mu} =0$.\footnote{In the Lagrangian approach to 
noncommutative QFT, $\theta^{0 \mu} \neq 0$ leads to unitarity violation. 
See \cite{BDFP02}, however, for a Hamiltonian approach in which this 
problem does not show up.}
But (\ref{Moyal-Rn}) is then obviously \emph{not} invariant under the usual 
action of the generators $P_\mu, M_{\mu\nu}$ of the Poincar{\'e} Lie algebra, 
which extends to functions via the derivation rule. In 
Hopf algebra language\footnote{Any Lie algebra $\mathfrak{g}$ can be 
turned into a Hopf algebra by first extending it to the universal 
enveloping algebra $\mathcal{U}(\mathfrak{g})$. Then  
$\Delta(Y) := Y \otimes 1 + 1 \otimes Y$ for 
$Y \in \mathcal{U}(\mathfrak{g})$ defines a homomorphism 
$\Delta : \mathcal{U}(\mathfrak{g}) \rightarrow \mathcal{U}(\mathfrak{g}) \otimes 
\mathcal{U}(\mathfrak{g})$, called the coproduct. Similarly, the antipode $S$ 
(generalized inverse) and the counit $\varepsilon$ are given by $S(Y) = -Y$ 
and $\varepsilon(Y)=0$, respectively. } 
the latter is given by 
$Y \triangleright (f h) \equiv Y \triangleright \mathbf{m}(f \otimes h)
  = \mathbf{m} \circ \Delta(Y) \triangleright (f \otimes h)$, 
using the coproduct $\Delta(Y)=Y \otimes 1 + 1 \otimes Y$.\footnote{Writing 
$\Delta(Y) = Y_{(1)} \otimes Y_{(2)}$ (Sweedler notation), we set 
$\Delta(Y) \triangleright (f \otimes h) := (Y_{(1)} \triangleright f) 
\otimes (Y_{(2)}\triangleright h)$. } 
Replacing the coproduct $\Delta$ with the \emph{twisted coproduct}  
$\Delta_{\mathcal{F}} = \mathcal{F} \Delta \mathcal{F}^{-1}$
(with $\mathcal{F}$ defined in (\ref{Moyal-prod})), then  
\be
   Y \triangleright ( f \star h) 
 = Y \triangleright \mathbf{m}_{\mathcal{F}}( f \otimes h)
 = \mathbf{m}_{\mathcal{F}} \circ \Delta_{\mathcal{F}}(Y) \triangleright 
   (f \otimes h)    \label{twPaction}
\ee
restores invariance \cite{CKNT04,Wess04,Szab06}. See also 
\cite{CPT05} for further implications.

\paragraph{Twisted (infinitesimal) diffeomorphisms and deformed gravity}
The above twist plays a crucial role in a recent formulation of 
Moyal-deformed differential geometry 
\cite{ABDMSW05,ADMW06,Wess06,Kurk+Sael06,Zupn07}. 
In classical differential geometry the action of an infinitesimal 
coordinate transformation generated by 
a vector field $\xi = \xi^\mu \pa_\mu$ on a scalar, vector and covector 
is given by
\be
   \delta_\xi \phi = - \xi \phi \, , \quad
   \delta_\xi V^\mu = - \xi V^\mu + (\pa_\nu \xi^\mu) V^\nu \, , \quad
   \delta_\xi a_\mu = - \xi a_\mu - (\pa_\mu \xi^\nu) a_\nu \, , 
   \label{tensor_transf}
\ee
respectively. In terms of the coproduct 
$\Delta(\delta_\xi) = \delta_\xi \otimes 1 + 1 \otimes \delta_\xi$, 
more general tensor transformation laws are recovered by applying 
$\delta_\xi \circ \mathbf{m} := \mathbf{m} \circ \Delta(\delta_\xi)$ 
to a tensor product. 

One can express the ordinary product of functions in terms of the star product 
via $f h = \mathbf{m}_{\mathcal{F}} \circ \mathcal{F} (f \otimes h) 
  = X_f \star h =: X_f \triangleright h$ with
\be
    X_f := \sum_{n=0}^\infty \frac{1}{n!} \left(-\frac{i}{2} \right)^n 
      \theta^{\mu_1\nu_1} \cdots \theta^{\mu_n\nu_n} (\pa_{\mu_1} \cdots 
      \pa_{\mu_n} f) \star \pa_{\nu_1} \cdots \pa_{\nu_n} \; . 
\ee
For a vector field $\xi$, we then have 
$\xi f = \xi^\mu \pa_\mu f = X_{\xi^\mu} \triangleright \pa_\mu f 
 =: X_\xi \triangleright f$ with the operator
\be
   X_\xi = \sum_{n=0}^\infty \frac{1}{n!} \left(-\frac{i}{2}\right)^n 
      \theta^{\mu_1\nu_1} \cdots \theta^{\mu_n\nu_n} (\pa_{\mu_1} \cdots 
      \pa_{\mu_n} \xi^\lambda) \star \pa_{\nu_1} \cdots \pa_{\nu_n} 
      \pa_\lambda 
\ee
(and correspondingly for a higher order differential operator replacing $\xi$). 
This yields a representation of the classical Lie algebra of vector fields, 
i.e. $[X_\xi , X_{\xi'} ]_\star = X_{[\xi,\xi']}$. 
Rewriting (\ref{tensor_transf}) as 
\be
   \hat{\delta}_\xi \phi = - X_\xi \triangleright \phi \, , \; 
   \hat{\delta}_\xi V^\mu = - X_\xi \triangleright V^\mu 
       + X_{\pa_\nu \xi^\mu} \triangleright V^\nu \, , \; 
   \hat{\delta}_\xi a_\mu = - X_\xi \triangleright a_\mu 
       - X_{\pa_\mu \xi^\nu} \triangleright a_\nu \, , 
\ee
and correspondingly for other tensors, one finds that
\be
  \hat{\delta}_\xi (S^{\mu_1 \ldots \mu_m}_{\nu_1\ldots\nu_n} 
    \star T^{\kappa_1 \ldots \kappa_r}_{\lambda_1\ldots\lambda_s} )
 = \mathbf{m}_{\mathcal{F}} \circ \Delta_{\mathcal{F}}(\hat{\delta}_\xi) 
    (S^{\mu_1 \ldots \mu_m}_{\nu_1\ldots\nu_n} 
    \otimes T^{\kappa_1 \ldots \kappa_r}_{\lambda_1\ldots\lambda_s} )
\ee
(which generalizes (\ref{twPaction})), hence the $\star$-product of 
tensors is again a tensor \cite{ABDMSW05}.
The way toward a noncommutative version of Einstein's equations is 
now straightforward. A covariant derivative can be introduced in analogy 
to its action on classical tensors, e.g. 
$D_\nu V^\mu = \pa_\nu \triangleright V^\mu 
+ \Gamma^\mu_{\lambda \nu} \star V^\lambda$. 
The connection $\Gamma^\mu_{\lambda \nu}$ has curvature 
$R^\kappa{}_{\lambda \mu \nu} 
   = \pa_\mu \triangleright \Gamma^\kappa_{\lambda \nu}
     - \pa_\nu \triangleright \Gamma^\kappa_{\lambda \mu}
     + \Gamma^\kappa_{\rho \mu} \star \Gamma^\rho_{\lambda \nu}
     - \Gamma^\kappa_{\rho \nu} \star \Gamma^\rho_{\lambda \mu}$ 
and Ricci tensor $R_{\mu\nu} = R^\kappa{}_{\mu \kappa \nu}$. 
A metric should be taken to be a symmetric and 
$\star$-invertible rank two tensor field $g_{\mu\nu}$, and we can 
impose vanishing torsion $\Gamma^\lambda_{\mu\nu} = \Gamma^\lambda_{\nu\mu}$ 
and metric compatibility $D_\lambda g_{\mu\nu} =0$. As in the 
classical case these conditions determine the connection in terms 
of the metric:
\be
   \Gamma^\lambda_{\mu\nu} = \frac{1}{2}( \pa_\mu \triangleright g_{\nu \kappa} 
      + \pa_\nu \triangleright g_{\mu \kappa} 
      - \pa_\kappa \triangleright g_{\mu \nu} ) \star g^{\kappa \lambda} \, , 
\ee
where $g^{\kappa \lambda}$ is the $\star$-inverse of $g_{\mu \nu}$. 
Because of noncommutativity there 
are two curvature scalars: $R = g^{\mu\nu} \star R_{\nu \mu}$ and 
 $R' = R_{\nu \mu} \star g^{\mu\nu}$. We refer to \cite{ABDMSW05} 
for further details and the construction of a deformed Einstein-Hilbert 
action functional. 
One can then ask whether this structure shows up in, say, string theory. 
This seems not to be the case. The above twisted gravity theory does 
not match the dynamics of closed strings in a constant $B$-field 
(beyond Seiberg-Witten approximation) \cite{AMV06}. 
What we probably should more worry about is the fact that the above 
formalism apparently distinguishes a class of coordinate systems, namely 
that with respect to which the twist operator $\mathcal{F}$ is defined 
(see also \cite{CTZZ06}).

\paragraph{Yet some other approaches}
Keeping $\theta^{\mu\nu}$ constant, and noting that  
$[x^\mu , f(x)]_\star = i \, \theta^{\mu\rho} \pa_{\rho} f(x)$, 
we find that an infinitesimal coordinate transformation 
$x^{\mu'} = x^\mu + \xi^\mu(x)$ leaves (\ref{Moyal-Rn}) invariant 
if $\theta^{\rho [\mu} \pa_{\rho} \xi^{\nu]}=0$. 
This is solved by $\xi^\mu = \theta^{\mu\nu} \pa_\nu f$ with a 
function $f(x)$. With respect to this \emph{restricted} class of coordinate 
transformations, one can then develop a formalism of geometry and 
General Relativity \cite{Calm+Koba05,Calm+Koba06}.  

The standard setup of a gauge theory on a noncommutative space 
requires that the anticommutator of Lie algebra elements lies 
in the Lie algebra\footnote{If $X,Y$ are elements of a Lie algebra and 
$a,b$ elements of some algebra $\A$, then 
$[a X , b Y] = \frac{1}{2} \{a,b\} [X,Y] + \frac{1}{2} [a,b] \{ X,Y \}$. 
For this to be Lie-algebra-valued, either $\A$ has to be commutative 
or the anticommutator $\{ X,Y \}$ has to lie in the Lie algebra.}, 
which is the case for a general linear group or $u(N)$ in the fundamental 
representation. A ``gravity theory'' formulated as a gauge theory 
on the Moyal space-time is thus necessarily complexified, see 
\cite{Cham00,Cham01,Cham03,Cham04,Moff00,Card+Zano03}. The problems 
with diffeomorphism invariance are still present, of course.

\section{DSR and $\kappa$-Poincar{\'e} symmetry}
In ``Doubly Special Relativity'' (DSR) (see \cite{Amel01,Kowa04}, for instance) 
the basic postulate is that 
all inertial observers should not only agree about the value of the speed of 
light, but also on the value of the Planck length $\ell_P$. 
A realization of this idea obviously requires a deformation of the Poincar{\'e} 
symmetry of Special Relativity. This can be achieved by stepping beyond 
the classical notion of symmetry toward the generalization offered by 
Hopf algebras. Indeed, a realization is the 
so-called $\kappa$-Poincar{\'e} algebra \cite{LRNT91,Maji+Rueg94} 
with $\kappa = \ell_P^{-1}$. 
This is one of the weakest Hopf algebra deformations of the Poincar{\'e} Lie 
algebra.\footnote{Hopf algebras comprise generalizations of Lie algebras 
as well as Lie groups. Correspondingly, there is also a $\kappa$-deformation 
of the Poincar{\'e} \emph{group} \cite{Zakr94} 
(as a ``matrix quantum group'' \cite{Woro87}).}
It is given by \cite{Maji+Rueg94}
\begin{eqnarray}
 && [ P_\mu , P_\nu ] = 0 \, , \quad
    [ N_i , P_0 ] = P_i \, , \quad
    [N_i , P_j] = \delta_{ij} \, \Big( \frac{\kappa}{2} (1-e^{-2P_0/\kappa})
                  + \frac{1}{2\kappa} \vec{P}^2 \Big) 
                  - \frac{1}{\kappa} \, P_i P_j  \nonumber \\
  && [N_i , N_j] = - \epsilon_{ijk} \, M_k \, , \quad
     [M_i , M_j] = \epsilon_{ijk} \, M_k \, , \quad
     [M_i , N_j] = \epsilon_{ijk} \, N_k   \nonumber  \\
  && [M_i , P_j] = \epsilon_{ijk} \, P_k \, , \quad 
     [M_i , P_0] = 0 \, , \quad 
     [N_i , P_0] = P_i    \label{4d-kappa-P}
\end{eqnarray}
(where $\mu,\nu=0,1,2,3$ and $i,j=1,2,3$), and  
\begin{eqnarray}
  && \Delta(P_0) = P_0 \otimes 1 + 1 \otimes P_0 \, , \quad
     \Delta(P_i) = P_i \otimes 1 + e^{-P_0/\kappa} \otimes P_i \, , 
              \nonumber \\
  && \Delta(M_i) = M_i \otimes 1 + 1 \otimes M_i \, , \quad
     \Delta(N_i) = N_i \otimes 1 + e^{-P_0/\kappa} \otimes N_i  
                 + \kappa^{-1} \, \epsilon_{ijk} \, P_j \otimes M_k 
      \; . \qquad
\end{eqnarray} 
In this (``bicrossproduct'' \cite{Maji+Rueg94}) formulation, the Lorentz 
sector is precisely that of the Poincar{\'e} algebra, and 
\begin{equation}
  2 \, \kappa^2 \cosh(P_0/\kappa) - \vec{P}^2 e^{P_0/\kappa} 
  = 2 \, \kappa^2 + P_0^2 - \vec{P}^2 - \kappa^{-1} \vec{P}^2 P_0
    + \mathcal{O}(\kappa^{-2}) 
\end{equation}
lies in the center of the algebra (i.e. commutes with all elements), thus 
has a fixed value in an irreducible representation. This leads to nonlinear corrections to the classical energy-momentum relations. Unlike the case 
of a Lie algebra, where only \emph{linear} transformations of the generators 
are allowed, there is now 
a huge freedom of \emph{nonlinear} transformations (also involving $\kappa$), 
even if we demand that the limit $\kappa \to 0$ reproduces the standard generators 
of the Poincar{\'e} group. Some additional input (to be expected from a 
quantum gravity theory), is thus needed to determine the ``physical'' 
energy and momentum. 

The coproduct is a rule to compose representations and thus to build 
multi-particle systems. Because of its asymmetry it appears to be 
difficult to make physical sense of the results in case of the 
$\kappa$-Poincar{\'e} algebra.
In any case, $\kappa$-Poincar{\'e} is an interesting example from 
which we can learn about generalized symmetries (quantum groups) 
in a physical context. 
Although quite a lot has been published about DSR, it has by far not 
reached the status of a physical theory as compared with SR.

\paragraph{$\kappa$-Poincar{\'e} from three-dimensional quantum gravity}
The Lie algebra of the Poincar\'e group in three space-time dimensions 
is given by\footnote{This is obtained from \eqref{4d-kappa-P}, reduced to 
$2+1$ dimensions (in which case there is only a single rotation generator  
$M := J_0$, such that $[ M , N_i ] = \epsilon_{ij} \, N^j$, 
$[ M , P_0 ] = 0$, $[ M , P_i ] = \epsilon_{ij} \, P^j$), 
in the limit as $\kappa \to \infty$. We have to identify $N_i = J_{i0}$, 
$i=1,2$. Indices are shifted with $\eta = \mbox{diag}(-1,1,1)$
and we use $\epsilon_{012}=1$. } 
\begin{eqnarray}
 \lbrack J_a,J_b\rbrack = \epsilon_{abc}\,J^c \, , \quad
 \lbrack J_a , P_b \rbrack = \epsilon_{abc}\,P^c \, , \quad
 \lbrack P_a,P_b \rbrack = 0 \, ,
\end{eqnarray}
where $J^c = {1\over 2} \, \epsilon^{c a b} \, J_{ab}$, and 
$J_{ab}$ are generators of $SO(2,1)$. 
Replacing $\lbrack P_a,P_b \rbrack = 0$ by
\begin{equation}
   \lbrack P_a,P_b \rbrack = \mp \ell^{-2} \, \epsilon_{abc} \, J^c 
\end{equation}
where $\ell$ is a parameter with dimension of length, we have for ``-'' 
the Lie algebra of $SO(3,1)$ and for ``+'' that of $SO(2,2)$. 
A (dimensionless) connection $A = \omega^a \, J_a + \theta^a \, P_a$ 
then has the field strength
\begin{eqnarray}
 F = \d A + A \wedge A
   = ( \mathcal{R}^c \mp \frac{1}{2 \ell^2} \epsilon^c{}_{ab} \, 
     \theta^a \wedge \theta^b ) \, J_c + \Theta^c \, P_c  \, ,
\end{eqnarray}
with curvature and torsion 
\begin{equation}
   \mathcal{R}^c := \d \omega^c + {1\over 2} \, \epsilon^c{}_{ab} \,
       \omega^a \wedge \omega^b 
      =: \frac{1}{2} \,\epsilon^{cab}\,\mathcal{R}_{ab} \, , \quad
   \Theta^c := \d \theta^c + \epsilon^c{}_{ab} \, \omega^a \wedge 
     \theta^b  \; .
\end{equation}
Using the invariant inner product given by 
$\langle J_a ,P_b \rangle = \ell^{-1} \eta_{ab}$, $\langle J_a,J_b \rangle = 0
 = \langle P_a,P_b \rangle$, one can construct a (dimensionless) Chern-Simons form 
\cite{Achu+Town86,Witt88,Carl05Liv}:
\begin{eqnarray}
   \langle A \wedge \d A + \frac{2}{3} \, A \wedge A \wedge A \rangle 
  = \ell^{-1} (\mathcal{R}^{ab} \mp \frac{1}{3 \ell^2} \theta^a 
    \wedge \theta^b ) \wedge \epsilon_{abc} \, \theta^c 
     -\d (\ell^{-1} \omega^c \wedge \theta_c) \, , 
\end{eqnarray}
which, up to an exact form, is the three-dimensional Einstein-Cartan 
Lagrangian with cosmological constant $\Lambda = \pm \ell^{-2}$, if 
we identify $k = \ell/\ell_P$.\footnote{If 
$\theta^a$, $a=0,1,2$, form a coframe (``dreibein''), then 
$g = \eta_{ab} \theta^a \otimes \theta^b$ defines a metric. But here $\theta^a$ 
is a gauge potential which in general does not constitute a coframe. 
The field equations $F=0 \; \Leftrightarrow \; \{ \, \Theta^c =0 \mbox{ and } \mathcal{R}^c=0 \, \}$ even admit exact solutions with $\theta^a=0$. 
Allowing a ``degenerate dreibein'' is crucial for treating 
three-dimensional gravity as a Chern-Simons gauge theory, but it means 
a serious departure from the usual understanding of gravity, see 
also \cite{Witt07}. } 
\footnote{See also \cite{Cham90,MH90CS} for an analogous relation between 
higher-dimensional Chern-Simons and generalized gravity actions. 
Their moduli space is much more complicated than in three dimensions, however.} 
Hence
\begin{eqnarray}
   S_{\mathrm{CS}}
 = \frac{k}{4 \pi} \int \langle A \wedge \d A + \frac{2}{3} \, A^3 \rangle 
 = \frac{k}{4 \pi \, \ell} \int (R - 2 \Lambda) \sqrt{|\det(g_{ab})|} \, \d^3x 
\end{eqnarray}
modulo boundary terms.\footnote{Global definition 
of the Chern-Simons action and single-valuedness of $e^{i S_{\mathrm{CS}}}$ 
requires that the real constant $k$ has to be ``quantized'' \cite{Witt07}.} 

Choosing space-time as $\Sigma \times \mathbb{R}$ with a two-dimensional 
compact surface $\Sigma$ of genus $\mathrm{g}$, the Chern-Simons action 
fixes the rules of canonical quantization. 
The physical degrees of freedom are that of the moduli space of flat 
connections (i.e. the space of connections modulo gauge transformations). 
Considering holonomies of the connection along noncontractable loops, this space 
can be described as the space of homomorphisms from $\pi_1(\Sigma)$, 
the fundamental group of $\Sigma$, into the global gauge group $G$  
(which is $SO(3,1)$ or $SO(2,2)$) \cite{Witt88,NRZ90}. 
For $U,V \in G$ representing two intersecting loops, one can define 
invariants, for which the quantization implies commutation 
relations of the quantum deformation $\mathcal{U}_q(so(3,1))$ of (the universal 
enveloping algebra of) $so(3,1)$, respectively $\mathcal{U}_q(so(2,2))$, 
where $q$ is a certain function of $k$. 
For positive cosmological constant, 
one obtains $\ln(q) \ell \approx \ell_P = \kappa^{-1}$ for small 
$\ell_P/\ell$ \cite{ASS04}. 
On account of this relation, the limit $\ell \to \infty$ 
(i.e. $\Lambda \to 0$) maps $\mathcal{U}_q(so(3,1))$ to the 
$\kappa$-Poincar{\'e} algebra, and correspondingly for 
$\mathcal{U}_q(so(2,2))$.
An essential ingredient in the derivation of this result is the 
nontrivial holonomy caused by nontrivial topology of the surface $\Sigma$, 
or ``punctures'' due to the presence of point particles. 
We refer to \cite{ASS04,FKS04} for further details, references, and 
also arguments toward similar results in the $3+1$-dimensional case, 
under special conditions.
In view of new insights \cite{Witt07} into the quantization of 
three-dimensional gravity, the above arguments may have to 
be reconsidered, however.

\paragraph{$\kappa$-Minkowski space}
This is the Hopf algebra with generators $x^\mu$ 
such that\footnote{We note that the nontrivial commutation relation of 
$\kappa$-Minkowski space is formally related to that of Klauder's 
``affine quantum gravity'' (see \cite{Klau06} and references therein). 
We recall the underlying idea. Starting from the canonical commutation 
relation $[q , p ] = i \,  \, \hbar \, I$, and multiplying by $q$,  
leads to the ``affine commutation relation'' 
$[ q , y ] = i \,  \, \hbar \, q$, where
$y = ( q p + p q )/2$. 
The news is now that, in contrast to the canonical commutation relation, 
the affine commutation relation allows that $q$ is selfadjoint with
\emph{positive} spectrum. Promoting $q$ to a spatial metric tensor, 
this would allow to respect metric positivity. }
$[x^i , x^0 ] = \kappa^{-1} \, x^i$, $[x^i , x^j ] = 0$, and 
$\Delta(x^\mu) = x^\mu \otimes 1 + 1 \otimes x^\mu$. 
The action of the momenta $P_\mu$ is, in the commutative case, given by 
the partial derivatives with respect to $x^\mu$. Because of the 
noncommutativity the rule is now
$P_\mu \triangleright :\!f(x^i,x^0)\!: \; = \; :\!\partial_\mu f(x^i,x^0)\!:$,
where $:\!f(x^i,x^0)\!:$ means ``normal ordering'': all powers of $x^0$ 
to the right. As a consequence of the above commutation relations, 
any analytic function of $x^0,x^i$ can be expressed 
as a sum of normal ordered functions. 
The further action of the $\kappa$-Poincar{\'e} algebra is 
given by 
$M_i \triangleright x^j = \epsilon_{ijk} \, x^k$, 
$M_i \triangleright x^0 = 0$,  
$N_i \triangleright x^j = -\delta_{ij} \, x^0$, 
$N_i \triangleright x^0 = - x^i$, 
and these definitions extend to the whole algebra via the 
familiar formula\footnote{For example, 
we have $P_0 \triangleright (fh) = (P_0 \triangleright f) \, h 
+ f \, (P_0 \triangleright h)$ and 
$P_i \triangleright (fh) = (P_i \triangleright f) \, h 
 + (e^{-P_0/\kappa} \triangleright f) \, (P_i \triangleright h)$.}
$Y \triangleright (fh) 
 = \mathbf{m} \circ \Delta(Y) \triangleright (f \otimes h) 
 = (Y_{(1)} \triangleright f) (Y_{(2)} \triangleright h)$  
for any element $Y$ of the $\kappa$-Poincar{\'e} algebra 
(cf. \cite{Maji+Rueg94}).\footnote{The 
reader should notice that we use the same symbol $\Delta$ for 
different coproducts.} 
It follows that $(x^0)^2 - \vec{x}^2 + 3 x^0/\kappa$ is 
invariant \cite{Maji+Rueg94}.

\section{Elements of Connes' NCG} 
In this section we sketch some of the main features and results of 
Connes' framework of ``spectral geometry'' 
(see in particular \cite{Conn95,GVF01,Schu01,CCM06,Conn+Marc07}).

\paragraph{Riemannian geometry in terms of the Dirac operator}
Let $M$ be an $n$-dimensional manifold with a pseudo-Riemannian metric 
$g = \eta_{ab} \, \theta^a \otimes \theta^b$, where $\theta^a$ is an
orthonormal coframe field. A Dirac spinor field on M has 
$\mathbb{C}^{2[n/2]}$-valued components 
$\psi$ with respect to $\theta^a$. 
With respect to another orthonormal coframe 
${\theta^a}' = L^a{}_b \, \theta^b$, 
related to the first by a function $L$ with values in the orthogonal group 
(invariance group of $\eta$), the components of the spinor field are 
$\psi' = S(L) \psi$, where $S$ is the representation of 
the orthogonal group determined by 
$S(L)^{-1} \, \gamma^a \, S(L) = L^a{}_b \, \gamma^b$ 
with constant matrices $\gamma^a$ satisfying the Clifford algebra relation 
$\gamma^a \gamma^b + \gamma^b \gamma^a = 2 \, \eta^{ab} \, I$.  
If $\omega^a{}_b$ are the Levi-Civita connection one-forms w.r. to $\theta^a$, 
we can introduce the covariant derivative and the Dirac operator
\be
     D_a \psi \; \theta^a := D \psi 
  := \d \psi + {1 \over 8} \, \omega_{ab} \, [ \gamma^a , \gamma^b ] \, \psi
     \, , \qquad
  \Dirac := \gamma^a D_a  \; .
\ee
In the Riemannian case (i.e. with a positive definite metric), 
the space of square-integrable spinor fields on $M$ with the inner product 
$(\psi, \chi) := \int_M \psi^\dagger \, \chi \, \sqrt{|\det(g_{ab})|} \, \d^n x$
provides us with a Hilbert space $\H$. 
Since $S$ is double-valued, more care is actually needed to define 
spinor fields. This leads to the notion of a \emph{spin$^c$ structure} and 
\emph{spin manifold} (see e.g. \cite{GVF01}). 

Connes observed that the geodesic distance on a Riemannian space can 
be recovered as follows from the Dirac operator (see \cite{GVF01} and 
references therein).
Let $M$ be a compact\footnote{If $M$ is not compact,
we should restrict $C^\infty(M)$ to functions which vanish sufficiently 
fast ``at infinity''.} 
spin manifold and $g$ a Riemannian metric. 
The geodesic distance $d(p,q)$ is then equal to
\be
   \mbox{dist}(p,q) := \sup \{ | p(f) - q(f) | \; ; \; 
     f \in C^\infty(M) , \; \| [\Dirac , f] \| \leq 1 \} \; . 
             \label{Connes_geod_dist}
\ee
Here we regard the points $p,q$ as pure states, so that $p(f) = f(p)$.
This suggests the following generalization:
\be
   \mbox{dist}(\phi,\phi') := \sup \{ | \phi(a) - \phi'(a) | \; ; \; a \in \A ,
   \; \| [ \mathcal{D} , a ] \| \leq 1 \} \, ,
\ee
where $\phi,\phi'$ are states\footnote{A state $\phi$ of a unital 
$C^\ast$-algebra is a normalized ($\phi(1)=1$) and positive 
($\phi(a^\ast a) \geq 0$) linear functional. It is ``pure'' if it is 
not a convex combination of other states. For a commutative algebra, 
pure states coincide with non-zero characters, i.e. homomorphisms into 
$\mathbb{C}$.}
of an algebra $\A$ of operators on a 
Hilbert space and $\mathcal{D}$ is a suitable analogue of the Dirac 
operator.\footnote{The expression 
$[ \mathcal{D} , a ]$ plays the role of a differential $\d a$. 
More generally, one-forms are given by 
$\sum_i a_i [ \mathcal{D} , b_i ]$ with $a_i, b_i \in \A$. }
The required structure is introduced next.

\paragraph{Spectral triples}
A \emph{spectral triple} 
$(\A,\H,\mathcal{D})$ consists of an involutive unital algebra $\A$, 
represented by bounded operators on a Hilbert space $\H$ (so that the 
antilinear involution ${}^\ast$ becomes the adjoint and the 
norm closure of the algebra is a $C^\ast$-algebra), 
and a selfadjoint operator $\mathcal{D}$ 
with compact\footnote{In order to address the case of a noncompact space, 
thus a non-unital algebra $\A$, one should require instead that the product 
of the resolvent with any element of $\A$ is a compact operator 
\cite{Conn95,GLMV02}.}
resolvent (hence the spectrum consists of countably many 
real eigenvalues) 
and such that $[\mathcal{D},a]$ is a bounded operator for each $a \in \A$. 

A spectral triple is called \emph{even} if $\H$ is endowed with a 
$\mathbb{Z}/2$-grading\footnote{This generalizes the chirality 
operator $\gamma_5$ of the ``commutative'' Dirac geometry in four dimensions.}, 
i.e. an operator $\gamma$ such that $\gamma = \gamma^\ast$, $\gamma^2 =1$, 
$[\gamma,a] =0$ for all $a \in \A$, and $\gamma$ anticommutes with $\mathcal{D}$. 

A spectral triple is called \emph{real of KO-dimension}\footnote{This 
is actually rather a signature than a ``dimension'' \cite{Barr07}. } 
$n \in \mathbb{Z}/8$ if there is an antilinear isometry 
(analogue of charge conjugation operator) $J : \H \rightarrow \H$ satisfying 
$J^2 = \varepsilon$, $J \mathcal{D} = \varepsilon' \mathcal{D} J$, and 
in the even case additionally $J \gamma = \varepsilon'' \gamma J$, where 
\begin{center}
\begin{tabular}{|c|rrrrrrrr|} \hline
$n$ & 0 & 1 & 2 & 3 & 4 & 5 & 6 & 7  \\
\hline
$\varepsilon$ & 1 & 1 & -1 & -1 & -1 & -1 & 1 & 1 \\
$\varepsilon'$ & 1 & -1 & 1 & 1 & 1 & -1 & 1 & 1 \\
$\varepsilon''$ & 1 &  & -1 &  & 1 &  & -1 &  \\ 
\hline
\end{tabular}
\end{center}
Moreover, $[a, J b^\ast J^{-1}]=0$ and 
$[[\mathcal{D},a],J b^\ast J^{-1}] =0$ for all $a,b \in \A$.\footnote{The 
first condition has its origin in Tomita-Takesaki theory (cf. \cite{Conn07}) 
and the second generalizes the property of the classical Dirac operator to 
be a first order differential operator.} 

Any compact Riemannian spin manifold $M$ gives 
rise to a real spectral triple of KO-dimension $n=\mathrm{dim}(M)$ mod $8$ 
with $\A = C^\infty(M)$. 
Conversely, given a real spectral triple with a \emph{commutative} 
unital algebra, such that certain additional conditions hold 
(which we do not list here), 
a compact Riemannian spin manifold can be constructed from it 
\cite{Renn+Vari06}. Up to unitary equivalence and 
spin structure preserving diffeomorphisms, compact Riemannian spin 
manifolds are in one-to-one correspondence with ``commutative'' 
real spectral triples subject to the aforementioned additional conditions. 

It should be noticed, however, that a Riemannian space cannot be 
reconstructed from the knowledge of the \emph{spectrum} of its Dirac 
operator alone. There are non-isometric compact Riemannian spin manifolds 
with Dirac operators having the same spectrum.

\paragraph{Spectral action}
As an analogue of the Einstein action in terms of the Dirac operator, 
Connes and Chamseddine \cite{Cham+Conn97,CCM06} proposed the 
{\em spectral action}
\be
    S(\Dirac,m) := \mathrm{Tr} f(\Dirac/m) \, , 
\ee
where $m$ is a parameter such that $\Dirac/m$ is dimensionless, 
and $f$ a positive even function chosen such that the trace exists. 
Via the heat kernel expansion method\footnote{A good review is \cite{Vass03}. 
See also \cite{Vass07} for heat kernel expansion of the spectral action on 
some noncommutative spaces like Moyal plane and noncommutative torus.}, 
in four dimensions ($n=4$) one obtains \cite{CCM06}\footnote{
Our convention for the Riemann tensor $R^{\kappa}{}_{\lambda\mu\nu} 
= \pa_\mu \Gamma^\kappa_{\lambda \nu} - \ldots$ differs by a minus sign 
from that e.g. in \cite{CCM06,Conn+Marc07}. }
\be
   S(\Dirac,m) &=& \frac{1}{16 \pi \mathcal{G}} \int_M (-R + 2 \Lambda) \, 
   \sqrt{\det(g)} \; d^4x  \nonumber \\
   && + \frac{f(0)}{10 \pi^2} \int_M ( \frac{11}{6} L_{GB} 
      - 3 C_{\mu\nu\kappa\lambda} C^{\mu\nu\kappa\lambda} ) \, \sqrt{\det(g)} 
      \; d^4x + \mathcal{O}(m^{-2})    \; ,
\ee
where $\mathcal{G} = \pi/(64 m^2 f_2)$, $\Lambda = 6 m^2 f_4/f_2$, 
$f_2 := \int_0^\infty v f(v) dv$, $f_4 := \int_0^\infty v^3 f(v) dv$. 
Furthermore, $L_{GB} = \frac{1}{4} \epsilon^{\mu\nu\kappa\lambda} 
\epsilon_{\alpha\beta\gamma\delta} R^{\alpha\beta}{}_{\mu\nu} 
R^{\gamma\delta}{}_{\kappa\lambda}$ is the Gauss-Bonnet term 
(which integrates to the Euler-Poincar{\'e} characteristic of $M$, up to 
some numerical factor) and 
$C^{\mu}{}_{\nu\kappa\lambda}$ is the Weyl (conformal) tensor of the metric. 
See \cite{Cham+Conn07} for additional boundary terms appearing in case of 
a manifold with boundary (with boundary conditions consistent with 
Hermiticity of the Dirac operator). 
The field theory action obtained from the spectral action 
has to be regarded as an \emph{effective} theory valid below the energy scale 
given by $m$.

\paragraph{Unification}
Kaluza-Klein theory attempted to unify all interactions by attaching 
an ``internal space'' to each space-time point, such that its isometries 
yield the gauge group of the standard model of elementary particle physics. 
In NCG the internal space should be replaced by 
an associative algebra $\A_i$ chosen in such a way that its group of 
(inner) automorphisms\footnote{An inner automorphism is determined by 
an invertible element $u \in \A_i$, which moreover has to be \emph{unitary} 
($u^\ast u = 1 = u u^\ast$), since an automorphism has to commute 
with the involution.}
coincides with this gauge group and it should possess 
a representation that reproduces the particle content of the 
standard model.\footnote{An advantage of using associative algebras 
instead of Lie algebras is the more constrained representation 
theory \cite{CCM06}. See also section~6.1 in \cite{Schu01}.}  
\bez
   \begin{array}{rcccl}
   \mbox{Diff}(M) \cong & \mathrm{Aut}(C^\infty(M)) & \times & \mathrm{Aut}(\A_i)
              & \cong U(1) \times SU(2) \times SU(3)  \\
      & \uparrow    &   & \uparrow &    \\
      & C^\infty(M) & \otimes & \A_i &    \\
      & \uparrow    &   & \uparrow &    \\
      & M           & \times & \mbox{``internal space''} &   
   \end{array}
\eez
Thus we have to extend the four-dimensional space-time algebra $C^\infty(M)$ 
to a larger algebra $\A = C^\infty(M) \otimes \A_i$ and find an appropriate 
real spectral triple, with a generalization $\mathcal{D}$ of the ordinary 
Dirac operator, such that the spectral action, extended by adding a fermionic 
part $\frac{1}{2} \langle J \psi , \mathcal{D} \psi \rangle$, reproduces 
the standard model action up to $\mathcal{O}(m^{-2})$. 
A good candidate for $\A_i$ is 
$\mathbb{C} \oplus \mathbb{H} \oplus M_3(\mathbb{C})$, where $\mathbb{H}$ 
are the quaternions and $M_3(\mathbb{C})$ the algebra of complex 
$3 \times 3$ matrices.\footnote{This has to be considered as 
a subalgebra of 
$\mathbb{C} \oplus \mathbb{H} \oplus \mathbb{H} \oplus M_3(\mathbb{C})$ 
\cite{CCM06,Conn+Marc07}. See also \cite{JKSS07} for a variant. }
A gauge field corresponding to the inner automorphisms 
of $\A_i$ can then be introduced by adding to the gravitationally 
coupled Dirac operator a term of the form $A + \varepsilon' JAJ^{-1}$ 
(which preserves the condition $J \mathcal{D} = \varepsilon' \mathcal{D} J$) 
with a self-adjoint one-form $A = \sum_i a_i [\mathcal{D},b_i]$. 
Another summand of the 
generalized Dirac operator corresponds to the fermion mass matrix. 
It turns out that the standard model coupled to gravity is obtained 
from a real spectral triple of KO-dimension $6$. 
We refer to \cite{CCM06,Conn+Marc07} for details and predictions. 
So far the focus is still on an ``understanding'' of the structure of the 
standard model of elementary particle physics in terms of (spectral) 
NCG, and the model had to be adapted \cite{Barr07,CCM06} 
to more recent findings of particle physics, like neutrino masses. 
$C^\infty(M) \otimes \A_i$ may well turn out to be a low energy  
approximation of some other noncommutative algebra.

\paragraph{Comments}
Connes' work includes a deep reformulation of \emph{Riemannian} geometry in 
terms of ``spectral geometry''. In many technical points it is 
restricted to positive definite metrics and their noncommutative analogues 
(see in particular \cite{Schu01} for some subtleties arising from 
the use of the Euclidean signature). 
Since a ``Wick rotation'' does not make sense for a general 
gravitational field, this Euclidean point of view cannot be satisfactory. 
Though ans\"atze toward a kind of pseudo-Riemannian version of Connes'
spectral Riemannian geometry have been proposed \cite{More03,Stro06,Pasc+Sita06}, 
a comparable reformulation of Lorentzian geometry, which after all 
is the physical one, is still out of sight. In particular, it appears 
to be impossible to define a Lorentzian analogue of the spectral 
action. 
Furthermore, the spectral action corresponds to a \emph{classical} field 
theory, it it not yet quantized. Parameters of the model are thus still 
subject to renormalization. 
See also \cite{Pasc07} for a critical account of Connes' NCG.

\section{Noncommutative differential geometry}
In classical differential geometry, the most basic geometric structure 
is given by a differentiable manifold, which is a topological space 
equipped with a ``differential structure''. The latter allows to define 
vector fields (sections of the tangent bundle), and then 
differential one-forms are introduced as linear maps acting on vector fields. 
Since vector fields are derivations of the algebra of smooth functions 
on the manifold, one can think of generalizing them to derivations 
of an algebra \cite{Dubo88}. Though this works 
for some interesting examples (see \cite{Mado99} and references therein), 
there are other algebras which do not admit any nontrivial 
derivation.\footnote{For example, the algebra of functions on a finite set 
admits only the trivial derivation $\delta =0$. }
On the other hand, there is a universal generalization of the notion 
of differential forms.
\vskip.1cm

Let $\cal A$ be an associative algebra. 
A {\em differential calculus} $(\Omega,\d)$ over $\A$ consists of an 
$\mathbb{N}_0$-graded algebra $\Omega = \bigoplus_{r\geq 0} \Omega^r$ 
with $\A$-bimodules $\Omega^r$, $\Omega^0=\A$, and a linear map
$ \d \, : \,  \Omega^r \rightarrow \Omega^{r+1}$ with the properties
\begin{eqnarray}
  \d^2 = 0 \, , \qquad
  \d (\alpha \, \beta) = (\d \alpha) \, \beta + (-1)^r \, \alpha \, \d \beta  
     \quad \mbox{(Leibniz rule)}
\end{eqnarray}
where $\alpha \in \Omega^r$ and $\beta \in \Omega$. 
\vskip.1cm

There are, however, \emph{many} differential calculi associated with a given 
algebra $\A$, the biggest being the ``universal differential envelope''. 
What is their significance? If $\A$ is the algebra of functions on a 
discrete set $M$, there is 
a bijective correspondence between (first order) differential calculi 
and digraphs on $M$ (so that the elements of $M$ are the vertices of the 
directed graph) \cite{DMH94graph}. An arrow from one point to another 
represents a discrete partial derivative component of the exterior 
derivative $\d$ in this ``direction''. A special example is the oriented 
hypercubic lattice digraph underlying lattice gauge theory \cite{DMHS93b,DMHS93a}. 
Thus, in the case of a discrete set, the choice of a differential calculus 
determines which points are neighbors. Typically no such interpretation 
exists in case of a calculus on a noncommutative algebra. 
The choice of a calculus has to be made according to the application 
one has in mind. Here are some possibilities to select certain calculi:
\begin{itemize}
\item A differential calculus can be defined in terms of a more basic structure. 
In Connes' NCG this is done via a generalized Dirac operator. 
\item If the algebra admits symmetries, these can be imposed on the calculus. 
Examples are bicovariant differential calculi on quantum groups 
\cite{Woro89,Klim+Schm97}.
\item Demanding the existence of a ``classical basis'' $\theta^i$ of one-forms: 
$\theta^i \, a = a \, \theta^i$ for all $a \in \A$ \cite{Mado99,BGMZ06}. 
In many cases there exists an ``almost classical basis'': 
$\theta^i \, a = \phi_i(a) \, \theta^i$ with automorphisms $\phi_i$ of $\A$ 
\cite{DMH04auto,DMH04CJP}. 
\end{itemize} 

As ``diffeomorphism group'' of the ``generalized manifold'' $(\A, \Omega)$ 
we should regard the automorphism group $\mathrm{Aut}(\Omega) \subset \mathrm{Aut}(\A)$. 

Further geometric notions
can be built on top of a differential calculus (and will depend on 
its choice, of course). 
In the algebraic language, ``fields'' on a manifold, or sections of a 
vector bundle, generalize to elements of a left (or right) $\A$-module 
$\mathcal{M}$.

A \emph{connection} on $\mathcal{M}$ 
(here we consider a \emph{left} $\A$-module) is a linear map
$\nabla \, : \, \mathcal{M} \rightarrow \Omega^1 \otimes_{\A} \mathcal{M}$ 
such that
\be
    \nabla (f \, \psi) = \d f \otimes_{\A} \psi + f \, \nabla \psi
\ee
for $f \in \A$ and $\psi \in \mathcal{M}$. It extends to a linear map
$\nabla \, : \; \Omega \otimes_{\A} \mathcal{M} \rightarrow \Omega 
\otimes_{\A} \mathcal{M}$ via
\be
   \nabla (\alpha \otimes_{\A} \psi) = \d \alpha \otimes_{\A} \psi 
     + (-1)^r \, \alpha \, \nabla \psi
   \qquad  \alpha \in \Omega^r, \; \psi \in \mathcal{M}  \; .
\ee
The \emph{field strenth}, or {\em curvature}, of the connection $\nabla$ 
is the map ${\cal R} = - \nabla^2$. 

If $\mathcal{M} = \Omega^1$, the connection $\nabla$ is a \emph{linear connection} 
with \emph{torsion} $\Theta = \d \circ \pi - \pi \circ \d \, : \, 
\Omega \otimes_{\A} \Omega^1 \rightarrow \Omega$, where 
$\pi$ is the projection $\Omega \otimes_{\A} \Omega^1 \rightarrow \Omega$. 

These are quite natural and universal definitions. 
If we had also a suitable concept of a \emph{metric} at hand, we could  
formulate a generalization of Einstein's equations on a 
``generalized manifold'' $(\A, \Omega)$. In particular this would allow 
to explore deformations of the classical Einstein equations (in general 
without a concrete expectation of what we could gain in this way). 
Among the various ways to introduce mathematically\footnote{An interpretation 
in terms of physical measurements has to follow.} 
a concept of a metric in NCG are the following. 
\begin{itemize}
\item In Connes' approach a (generalized) \emph{Riemannian} metric is 
defined in terms of a (generalized) Dirac operator and the spectral 
action generalizes the Einstein-Hilbert action. But all this is 
essentially bound to the Euclidean regime. 
\item The algebraic approach suggests to define a metric as a map
$g : \Omega^1 \otimes_\A \Omega^1 \rightarrow \A$ (see e.g. \cite{Mado99} 
for some examples), or an element $g \in \Omega^1 \otimes_\A \Omega^1$ 
(with suitable reality and invertibility properties). We mention now  
that this is not always appropriate.
\item In a formulation of pseudo-Riemannian geometry on discrete sets 
\cite{DMH99dRg,DMH03grpl1,DMH03grpl2} the correct geometric interpretation 
requires $g = g_{\mu \nu} \, \d x^\mu \otimes_L \d x^\nu$, 
where $\otimes_L$ is the \emph{left-linear tensor product}, which satisfies 
$(f \, \alpha) \otimes_L (h \, \beta) = f \, h \, \alpha \otimes_L \beta$ 
for $f,h \in \A$ and $\alpha, \beta \in \Omega^1$.\footnote{The left-linear 
tensor product does \emph{not} exist for a \emph{non}commutative algebra $\A$.} 
\item If a differential calculus possesses a ``classical basis'' $\theta^i$ 
(see above), one may postulate it to be orthonormal and introduce in this way 
a metric $g = \eta_{i j} \, \theta^i \otimes_\A \theta^j$ (see e.g. 
\cite{BGMZ06}). Note that $g$ has the left-linearity property in this basis. 
\end{itemize}
In particular, a Lorentzian signature can be implemented. We refer 
to the references cited above for further details of special approaches 
(and further obstacles to build a deformation or noncommutative analogue 
of Einstein's theory). 

If $\A$ is a deformation of a commutative algebra, say the algebra of (smooth) 
functions on $\mathbb{R}^n$, there may exist differential calculi over $\A$ 
which do \emph{not} tend to the classical calculus of differential forms when 
the deformation vanishes (see \cite{MH+Reut93,DMH93stoch} for an example). 
In the following subsection we consider a class of such ``noncommutative 
differential calculi'' on $\mathbb{R}^n$ and show how a metric can emerge 
from it.

\subsection{A class of noncommutative differential calculi on $\mathbb{R}^n$}
Let $\A$ be the algebra of functions generated by commuting objects $x^\mu$, 
$\mu = 1,\ldots,n$, e.g. coordinate functions on $\mathbb{R}^n$. 
A class of differential calculi is then determined by
\be
   [ \d x^\mu, x^\nu ] = \ell \, C^{\mu \nu}{}_\kappa \, \d x^\kappa \, , 
        \label{dc-C}
\ee
where $\ell$ is a constant with dimension of length and $C^{\mu \nu}{}_\kappa$ 
are dimensionless functions of the coordinates, which have to satisfy the 
conditions $C^{\mu \nu}{}_\kappa = C^{\nu \mu}{}_\kappa$ and 
$C^{\mu\kappa}{}_\lambda \, C^{\nu \lambda}{}_\kappa 
   = C^{\nu\kappa}{}_\lambda \, C^{\mu \lambda}{}_\kappa$ 
\cite{DMHS93a,BDMH95,DMH00CJP}.\footnote{In terms of the matrices $C^\mu$ 
with entries $(C^\mu)^\nu{}_\kappa = C^{\mu \nu}{}_\kappa$, the last 
condition means that they have to commute. The two conditions 
imply that $x^\mu \bullet x^\nu := C^{\mu \nu}{}_\kappa \, x^\kappa$ 
determines a commutative and associative product. Such algebras 
play a role in a description of topological field theories as 
lattice models \cite{Bach+Petr93,FHK94}. 
}
Thinking of a space-time model, a natural candidate for $\ell$ 
would be the Planck length $\ell_P$.
The above deformation of the classical differential calculus then modifies 
the kinematical structure of space-time at the Planck scale. 

We assume that $\{ \d x^\mu \}$ is a basis of $\Omega^1$ as a left- and as a right $\A$-module. Generalized partial (left- and right-) derivatives can then 
be introduced via
\be
   \d f = (\pa_{+\mu} f) \, \d x^\mu = \d x^\mu \, (\pa_{-\mu} f)  \; .
\ee
The concrete form of the generalized partial derivatives depends 
on the structure functions $C^{\mu\nu}{}_\kappa$. A ``coordinate transformation'' 
(diffeomorphism) should now be an invertible map $x^\mu \mapsto x^{\mu'}(x^\nu)$ 
with the property that $\pa_{+\nu} x^{\mu'}$ is invertible. 
This allows to generalize the notions of manifold and tensors. We find
\be
     [ \d x^{\mu'}, x^{\nu'} ] 
 &=& \pa_{+\kappa} x^{\mu'} \, [ \d x^{\kappa}, x^{\nu'} ]
  =  \pa_{+\kappa} x^{\mu'} \, [ \d x^{\nu'}, x^\kappa ]     \nonumber \\
 &=& \pa_{+\kappa} x^{\mu'} \, \pa_{+\lambda} x^{\nu'} \, [ \d x^\lambda, x^\kappa ]
  = \ell \, \pa_{+\kappa} x^{\mu'} \, \pa_{+\lambda} x^{\nu'} \, 
    C^{\kappa \lambda}{}_\sigma \, \d x^\sigma \, , 
\ee 
using the commutativity of $\A$ and the derivation property of $\d$. 
Comparison with (\ref{dc-C}) implies 
$ C^{\mu' \nu'}{}_{\kappa'} 
 = \pa_{+\kappa} x^{\mu'} \, \pa_{+\lambda} x^{\nu'} \, 
    C^{\kappa \lambda}{}_\sigma \, \pa_{+\kappa'} x^{\sigma}$. 
As a consequence,\footnote{This can be written as 
$g^{\mu \nu} = \mathrm{tr}(C^\mu C^\nu)$. }
\be
   g^{\mu \nu} := C^{\mu \kappa}{}_\lambda \, C^{\lambda \nu}{}_\kappa 
         \label{C->g}
\ee
(see also \cite{FHK94}) is symmetric and obeys the tensor transformation law 
$g^{\mu' \nu'} = \pa_{+\kappa} x^{\mu'} \, \pa_{+\lambda} x^{\nu'} \,
                   g^{\kappa \lambda} $. 
If an inverse $g_{\mu \nu}$ exists\footnote{This is the case iff the 
algebra determined by the $C^{\mu\nu}{}_\kappa$ is semi-simple \cite{FHK94}.}, 
then it is also a tensor, i.e. 
$g_{\mu' \nu'} = \pa_{+\mu'} x^\kappa \, \pa_{+\nu'} x^\lambda \,
 g_{\kappa \lambda} $. 
\vskip.1cm
\noindent
\emph{Example~1.} 
If there are coordinates such that $C^{\mu\nu}{}_\kappa = \delta^\mu_\kappa 
\, \delta^\nu_\kappa$, then 
$[ \d x^\mu, x^\nu ] = \ell \, \delta^{\mu\nu} \, \d x^\nu$, which is the 
hypercubic lattice differential calculus \cite{DMHS93b,DMHS93a}. 
In this case the generalized partial derivatives are the left/right 
discrete derivatives on a lattice with lattice spacing $\ell$, 
and we have the Euclidean metric $g_{\mu \nu} = \delta_{\mu\nu}$, 
as expected.\footnote{With a slight modification one obtains the Minkowski metric.} 
\vskip.1cm
\noindent
\emph{Example~2.} Let $\gamma^{\mu\nu}$ be components of a symmetric 
tensor field and $\tau = \tau_\mu \d x^\mu$ a one-form on a manifold, 
such that $\gamma^{\mu\nu} \tau_\nu =0$  (``generalized Galilei structure''). 
Then (\ref{dc-C}) with $C^{\mu\nu}{}_\kappa = \gamma^{\mu\nu} \tau_\kappa$ 
is invariant under general coordinate transformations and thus extends 
to the whole manifold. In this case the tensor (\ref{C->g}) vanishes. 
We refer to \cite{DMH92_grav,MH+Reut93,DMH93stoch,Dima+Tzan96} for appearances 
of this structure, which in particular makes contact with stochastic 
calculus on manifolds. See also \cite{Kauf04} for related work. 
\vskip.1cm

In the limit $\ell \to 0$, where the differential calculus (\ref{dc-C}) 
becomes the ``classical'' one, we should expect that the generalized 
partial derivatives become ordinary partial derivatives (when acting 
on smooth functions). 
In this limit the metric decouples from the differential structure, whereas 
for $\ell \neq 0$ it is a property of the differential calculus.

\section{Final remarks}
We have briefly reviewed a variety of ideas about 
``noncommutative space-time'' and some ans\"atze toward 
corresponding generalizations of General Relativity. 
Among the most interesting developments is certainly the reformulation of 
the whole standard model of elementary particle physics including gravity 
in a concise NCG language by Connes and his disciples. 
This primarily aims at a better understanding 
of the quite complicated structure of the standard model. 
Since elementary particle physics is what tells us about the 
small scale structure of space-time, this is a promising route 
toward a deeper unification of space-time, particles and forces, 
though the lack of a Lorentzian version still presents a serious obstacle. 

In NCG a machinery similar to that of quantum physics is 
already introduced at a ``classical'' level. 
So there has to be another, apparently completely different level 
which introduces a similar machinery on top of the first. 
This appears to be a major complication and hardly satisfactory. 
We should rather hope that either both quantizations can be merged 
to a single one, or one induces the other automatically. 

Needless to say, there are many more interesting ideas and facts 
in the ``noncommutative world'' related to the notions of space-time 
and gravity than we touched upon in this short review. 
For some of them, in particular related to matrix models, we refer 
to \cite{Szab06}. 

\vskip.3cm
\noindent
\textbf{Acknowledgments.}
The author would like to thank the organizers, and especially Alfredo Macias, 
for the invitation to the III. Mexican Meeting on Mathematical and Experimental Physics.

\hyphenation{Post-Script Sprin-ger}


\begin{thebibliography}{100}

\bibitem{Gero72}
R.~Geroch,
 Einstein algebras,
 Commun. Math. Phys.
   26  (1972) 271--275.

\bibitem{Land+Marm91}
G.~Landi and G.~Marmo,
 Algebraic field theories, {E}instein algebras and noncommutative geometry,
 in: {\em General Relativity and Gravitational Physics}, eds. R.~Cianci,
  R.~de~Ritis, M.~Francaviglia, G.~Marmo, C.~Rubano, and P.~Scudellaro
 (World Scientific, Hong Kong, 1991)
   292.

\bibitem{Land97}
G.~Landi,
 {\em An {I}ntroduction to {N}oncommutative {S}paces and their {G}eometries},
 Vol.~51 of {\em Lecture Notes in Physics}
 (Springer, Berlin, 1997).

\bibitem{GVF01}
J.M. Gracia-Bondia, J.C. V{\'a}rilly, and H.~Figueroa,
 {\em Elements of {N}oncommutative {G}eometry}
 (Birh\"auser, Basel, 2001).

\bibitem{Corn+Schi02}
L.~Cornalba and R.~Schiappa,
 Nonassociative star product deformations for {$D$}-brane world-volumes in
  curved backgrounds,
 Commun. Math. Phys.
   225  (2002) 33--66.

\bibitem{Buch+Grun07}
D.~Buchholz and H.~Grundling,
 The resolvent algebra: a new approach to canonical quantum systems,
 arXiv:0705.1988.

\bibitem{BFFLS78a}
F.~Bayen, M.~Flato, C.~Fronsdal, A.~Lichnerowicz, and D.~Sternheimer,
 Deformation theory and quantization. {I}. {D}eformation of symplectic
  structures,
 Ann. Phys.
   111  (1978) 61--110.

\bibitem{BFFLS78b}
F.~Bayen, M.~Flato, C.~Fronsdal, A.~Lichnerowicz, and D.~Sternheimer,
 Deformation theory and quantization. {II}. {P}hysical applications,
 Ann. Phys.
   111  (1978) 111--151.

\bibitem{Groe46}
H.~J. Groenewold,
 On the principles of elementary quantum mechanics,
 Physica
   12  (1946) 405--460.

\bibitem{Moya49}
J.~E. Moyal,
 Quantum mechanics as a statistical theory,
 Proc. Cambridge Phil. Soc.
   45  (1949) 99--124.

\bibitem{Klim+Schm97}
A.~Klimyk and K.~Schm{\"u}dgen,
 {\em Quantum {G}roups and their {R}epresentations}
 (Springer, Berlin, 1997).

\bibitem{Snyd47}
H.S. Snyder,
 Quantized space-time,
 Phys. Rev.
   71  (1947) 38--41.

\bibitem{Yang47}
C.N. Yang,
 On quantized space-time,
 Phys. Rev.
   72  (1947) 874.

\bibitem{Mats+Well98}
H.-J. Matschull and M.~Welling,
 Quantum mechanics of a point particle in $(2+1)$-dimensional gravity,
 Class. Quantum Grav.
   15  (1998) 2981--3030.

\bibitem{Filk96}
T.~Filk,
 Divergencies in a field theory on quantum space,
 Phys. Lett. B
   376  (1996) 53--58.

\bibitem{Vari+Grac99}
J.C. V{\'a}rilly and J.M. Gracia-Bondia,
 On the ultraviolet behaviour of quantum fields over non-commutative manifolds,
 Int. J. Mod. Phys. A
   14  (1999) 1305--1323.

\bibitem{DMHS93b}
A.~Dimakis, F.~M\"uller-Hoissen, and T.~Striker,
 From continuum to lattice theory via deformation of the differential calculus,
 Phys. Lett. B
   300  (1993) 141--144.

\bibitem{DFR94}
S.~Dopplicher, K.~Fredenhagen, and J.~E. Roberts,
 Spacetime quantization induced by classical gravity,
 Phys. Lett. B
   331  (1994) 39--44.

\bibitem{DFR95}
S.~Dopplicher, K.~Fredenhagen, and J.~E. Roberts,
 The quantum structure of spacetime at the {P}lanck scale and quantum fields,
 Commun. Math. Phys.
   172  (1995) 187--220.

\bibitem{Gara95}
L.J. Garay,
 Quantum gravity and minimum length,
 Int. J. Mod. Phys. A
   10  (1995) 145--165.

\bibitem{Seib+Witt99}
N.~Seiberg and E.~Witten,
 String theory and noncommutative geometry,
 JHEP
   9909  (1999) 032.

\bibitem{Jack02}
R.~Jackiw,
 Physical instances of noncommuting coordinates,
 Nucl. Phys. Proc. Suppl.
   108  (2002) 30--36.

\bibitem{Hein+Ilde07}
T.~Heinzl and A.~Ilderton,
 Noncommutativity from spectral flow,
 arXiv:0704.3547.

\bibitem{Mado99ped}
J.~Madore,
 Noncommutative geometry for pedestrians,
 gr-qc/9906059.

\bibitem{Mado92}
J.~Madore,
 The fuzzy sphere,
 Class. Quantum Grav.
   9  (1992) 69--87.

\bibitem{Kont03}
M.~Kontsevich,
 Deformation quantization of {P}oisson manifolds,
 Lett. Math. Phys.
   66  (2003) 157--216.

\bibitem{Catt+Feld00}
A.S. Cattaneo and G.~Felder,
 A path integral approach to the {K}ontsevich quantization formula,
 Commun. Math. Phys.
   212  (2000) 591--611.

\bibitem{Szab06}
R.J. Szabo,
 Symmetry, gravity and noncommutativity,
 Class. Quantum Grav.
   23  (2006) R199--R242.

\bibitem{Cham01}
A.H. Chamseddine,
 Deforming Einstein's gravity,
 Phys. Lett. B
   504  (2001) 33--37.

\bibitem{CTZZ06}
M.~Chaichian, A.~Tureanu, R.B. Zhang, and X.~Zhang,
 Riemannian geometry of noncommutative surfaces,
 hep-th/0612128.

\bibitem{BDFP02}
D.~Bahns, S.~Doplicher, K.~Fredenhagen, and G.~Piacitelli,
 On the unitarity problem in space/time noncommutative theories,
 Phys. Lett. B
   533  (2002) 178--181.

\bibitem{CKNT04}
M.~Chaichian, P.P. Kulish, K.~Nishijima, and A.~Tureanu,
 On a Lorentz-invariant interpretation of noncommutative space–time and its
  implications on noncommutative QFT,
 Phys. Lett. B
   604  (2004) 98--102.

\bibitem{Wess04}
J.~Wess,
 Deformed coordinate spaces; derivatives,
 hep-th/0408080.

\bibitem{CPT05}
M.~Chaichian, P.~Pre\v{s}najder, and A.~Tureanu,
 New concept of relativistic invariance in noncommutative space-time: twisted
  {P}oincar{\'e} symmetry and its implications,
 Phys. Rev. Lett.
   94  (2005) 151602.

\bibitem{ABDMSW05}
P.~Aschieri, C.~Blohmann, M.~Dimitrijevi{\'c}, F.~Meyer, P.~Schupp, and
  J.~Wess,
 A gravity theory on noncommutative spaces,
 Class. Quantum Grav.
   22  (2005) 3511--3532.

\bibitem{ADMW06}
P.~Aschieri, M.~Dimitrijevi{\'c}, F.~Meyer, and J.~Wess,
 Noncommutative geometry and gravity,
 Class. Quantum Grav.
   23  (2006) 1883--1911.

\bibitem{Wess06}
J.~Wess,
 Einstein-{R}iemann gravity on deformed spaces,
 SIGMA
   2  (2006) 089.

\bibitem{Kurk+Sael06}
S.~K\"{u}rk\c{c}\"{u}o\v{g}lu and C.~S\"{a}mann,
 Drinfeld twist and general relativity with fuzzy spaces,
 hep-th/0606197.

\bibitem{Zupn07}
B.M. Zupnik,
 Reality in noncommutative gravity,
 Class. Quantum Grav.
   24  (2007) 15--26.

\bibitem{AMV06}
L.~{\'A}lvarez-Gaum{\'e}, F.~Meyer, and M.A. V{\'a}zquez-Mozo,
 Comments on noncommutative gravity,
 Nucl. Phys. B
   753  (2006) 92--117.

\bibitem{Calm+Koba05}
X.~Calmet and A.~Kobakhidze,
 Noncommutative general relativity,
 Phys. Rev. D
   72  (2005) 045010.

\bibitem{Calm+Koba06}
X.~Calmet and A.~Kobakhidze,
 Second order noncommutative corrections to gravity,
 Phys. Rev. D
   74  (2006) 047702.

\bibitem{Cham00}
A.H. Chamseddine,
 Complexified gravity in noncommutative spaces,
 hep-th/0005222.

\bibitem{Cham03}
A.H. Chamseddine,
 An invariant action for noncommutative gravity in four dimensions,
 J. Math. Phys.
   44  (2003) 2534--2541.

\bibitem{Cham04}
A.H. Chamseddine,
 {$SL(2,\mathbb{C}$} gravity with complex vierbein and its noncommutative
  extension,
 Phys. Rev. D
   69  (2004) 024015.

\bibitem{Moff00}
J.W. Moffat,
 Noncommutative quantum gravity,
 Phys. Lett. B
   491  (2000) 345--352.

\bibitem{Card+Zano03}
M.A. Cardella and D.~Zanon,
 Noncommutative deformation of four-dimensional {E}instein gravity,
 Class. Quantum Grav.
   20  (2003) L95--L103.

\bibitem{Amel01}
G.~Amelino-Camelia,
 Testable scenario for relativity with minimum length,
 Phys. Lett. B
   510  (2001) 255--263.

\bibitem{Kowa04}
J.~Kowalski-Glikman,
 Introduction to doubly special relativity,
 hep-th/0405273.

\bibitem{LRNT91}
J.~Lukierski, H.~Ruegg, A.~Nowicki, and V.N. Tolstoi,
 {$Q$} deformation of {P}oincar{'e} algebra,
 Phys. Lett. B
   264  (1991) 331--338.

\bibitem{Maji+Rueg94}
S.~Majid and H.~Ruegg,
 Bicrossproduct structure of $\kappa$-{P}oinca{\'e} group and non-commutative
  geometry,
 Phys. Lett. B
   334  (1994) 348--354.

\bibitem{Zakr94}
S.~Zakrzewski,
 Quantum {P}oincar{\'e} group related to the $\kappa$-{P}oincar{\'e} algebra,
 J. Phys. A: Math. Gen.
   27  (1994) 2075--2082.

\bibitem{Woro87}
S.L. Woronowicz,
 Compact matrix pseudogroups,
 Commun. Math. Phys.
   111  (1987) 613.

\bibitem{Achu+Town86}
A.~Ach\'ucarro and P.~Townsend,
 A {C}hern-{S}imons action for three-dimensional anti-de {S}itter supergravity
  theories,
 Phys. Lett. B
   180  (1986) 89--92.

\bibitem{Witt88}
E.~Witten,
 2+1 dimensional gravity as an exactly soluble system,
 Nucl. Phys. B
   311  (1988) 46--78.

\bibitem{Carl05Liv}
S.~Carlip,
 Quantum gravity in $2+1$ dimensions: {T}he case of a closed universe,
 Living Rev. Relativity
   8  (2005) 1--63.

\bibitem{Witt07}
E.~Witten,
 Three-dimensional gravity reconsidered,
 arXiv:0706.3359.

\bibitem{Cham90}
A.H. Chamseddine,
 Topological gravity and supergravity in various dimensions,
 Nucl. Phys. B
   346  (1990) 213--234.

\bibitem{MH90CS}
F.~M\"uller-Hoissen,
 From {C}hern-{S}imons to {G}auss-{B}onnet,
 Nucl. Phys. B
   346  (1990) 235--252.

\bibitem{NRZ90}
J.E. Nelson, T.~Regge, and F.~Zertuche,
 Homotopy groups and $(2+1)$-dimensional quantum de {S}itter gravity,
 Nucl. Phys. B
   339  (1990) 516--532.

\bibitem{ASS04}
G.~Amelino-Camelia, L.~Smolin, and A.~Starodubtsev,
 Quantum symmetry, the cosmological constant and {P}lanck-scale phenomenology,
 Class. Quantum Grav.
   21  (2004) 3095--3110.

\bibitem{FKS04}
L.~Freidel, J.~Kowalski-Glikman, and L.~Smolin,
 $2+1$ gravity and doubly special relativity,
 Phys. Rev. D
   69  (2004) 044001.

\bibitem{Klau06}
J.R. Klauder,
 Overview of Affine Quantum Gravity,
 Int. J. Geom. Meth. Mod. Phys.
   3  (2006) 81--94.

\bibitem{Conn95}
A.~Connes,
 Noncommutative geometry and reality,
 J. Math. Phys.
   36  (1995) 6194--6231.

\bibitem{Schu01}
T.~Sch\"ucker,
 Forces from {C}onnes' geometry,
 hep-th/0111236.

\bibitem{CCM06}
A.H. Chamseddine, A.~Connes, and M.~Marcolli,
 Gravity and the standard model with neutrino mixing,
 hep-th/0610241.

\bibitem{Conn+Marc07}
A.~Connes and M.~Marcolli,
 {\em Noncommutative {G}eometry, {Q}uantum {F}ields and {M}otives}, 
 ftp://ftp.alainconnes.org/bookjuly.pdf (2007).

\bibitem{GLMV02}
J.M. Gracia-Bondia, F.~Lizzi, G.~Marmo, and P.~Vitale,
 Infinitely many star products to play with,
 JHEP 04  (2002) 026.

\bibitem{Barr07}
J.W. Barrett,
 Lorentzian version of the noncommutative geometry of the standard model of
  particle physics,
 J. Math. Phys.
   48  (2007) 012303--1--7.

\bibitem{Conn07}
A.~Connes,
 On the fine structure of spacetime,
 ftp://ftp.alainconnes.org/shahnlong.pdf.

\bibitem{Renn+Vari06}
A.~Rennie and J.C. V\'{a}rilly,
 Reconstruction of manifolds in noncommutative geometry,
 math.QA/0610418.

\bibitem{Cham+Conn97}
A.H. Chamseddine and A.~Connes,
 The spectral action principle,
 Commun. Math. Phys.
   186  (1997) 731--750.

\bibitem{Vass03}
D.V. Vassilevich,
 Heat kernel expansion: user's manual,
 Phys. Rep.
   388  (2003) 279--360.

\bibitem{Vass07}
D.V. Vassilevich,
 Heat trace asymptotics on noncommutative spaces,
 arXiv:0708.4209.

\bibitem{Cham+Conn07}
A.H. Chamseddine and A.~Connes,
 Quantum gravity boundary terms from spectral action,
 arXiv:0705.1786.

\bibitem{JKSS07}
J.-H. Jureit, T.~Krajewski, T.~Sch\"ucker, and C.A. Stephan,
 On the noncommutative standard model,
 Acta Phys. Polon. B
   38  (2007) 3181--3202.

\bibitem{More03}
V.~Moretti,
 Aspects of noncommutative {L}orentzian geometry for globally hyperbolic
  spacetimes,
 Rev. Math. Phys.
   15  (2003) 1171--1217.

\bibitem{Stro06}
A.~Strohmaier,
 On noncommutative and pseudo-{R}iemannian geometry,
 J. Geom. Phys.
   56  (2006) 175--195.

\bibitem{Pasc+Sita06}
M.~Paschke and A.~Sitarz,
 Equivariant {L}orentzian spectral triples,
 math-ph/0611029.

\bibitem{Pasc07}
M.~Paschke,
 An essay on the spectral action and its relation to quantum gravity,
 in: {\em Quantum Gravity, Mathematical Models and Experimental Bounds}
 (Birkh\"auser, Basel, 2007)
   127--150.

\bibitem{Dubo88}
M.~Dubois-Violette,
 D{\'e}rivations et calcul diff{\'e}rentiel non commutatif,
 C. R. Acad. Sci. Paris, S{\'e}rie I
   307  (1988) 403--408.

\bibitem{Mado99}
J.~Madore,
 {\em Introduction to {N}on-commutative {G}eometry and its {P}hysical
  {A}pplications}
 (Cambridge University Press, Cambridge, 1999).

\bibitem{DMH94graph}
A.~Dimakis and F.~M\"uller-Hoissen,
 Discrete differential calculus, graphs, topologies and gauge theory,
 J. Math. Phys.
   35  (1994) 6703--6735.

\bibitem{DMHS93a}
A.~Dimakis, F.~M\"uller-Hoissen, and T.~Striker,
 Noncommutative differential calculus and lattice gauge theory,
 J. Phys. A: Math. Gen.
   26  (1993) 1927--1949.

\bibitem{Woro89}
S.L. Woronowicz,
 Differential calculus on compact matrix pseudogroups (quantum groups),
 Commun. Math. Phys.
   122  (1989) 125--170.

\bibitem{BGMZ06}
M.~Buri{\'c}, T.~Grammatikopoulos, J.~Madore, and G.~Zoupanos,
 Gravity and the structure of noncommutative algebras,
 hep-th/0603044.

\bibitem{DMH04auto}
A.~Dimakis and F.~M\"uller-Hoissen,
 Automorphisms of associative algebras and noncommutative geometry,
 J. Phys. A: Math. Gen.
   37  (2004) 2307--2330.

\bibitem{DMH04CJP}
A.~Dimakis and F.~M\"uller-Hoissen,
 Differential calculi on quantum spaces determined by automorphisms,
 Czech J. Phys.
   54  (2004) 1235--1241.

\bibitem{DMH99dRg}
A.~Dimakis and F.~M\"uller-Hoissen,
 Discrete {R}iemannian geometry,
 J. Math. Phys.
   40  (1999) 1518--1548.

\bibitem{DMH03grpl1}
A.~Dimakis and F.~M\"uller-Hoissen,
 Differential geometry of group lattices,
 J. Math. Phys.
   44  (2003) 1781--1821.

\bibitem{DMH03grpl2}
A.~Dimakis and F.~M\"uller-Hoissen,
 Riemannian geometry of bicovariant group lattices,
 J. Math. Phys.
   44  (2003) 4220--4259.

\bibitem{MH+Reut93}
F.~M\"uller-Hoissen and C.~Reuten,
 Bicovariant differential calculus on {$GL_{p,q}(2)$} and quantum subgroups,
 J. Phys. A: Math. Gen.
   26  (1993) 2955--2975.

\bibitem{DMH93stoch}
A.~Dimakis and F.~M\"uller-Hoissen,
 Stochastic differential calculus, the Moyal $\star$-product, and
  noncommutative geometry,
 Lett. Math. Phys.
   28  (1993) 123--137.

\bibitem{BDMH95}
H.~Baehr, A.~Dimakis, and F.~M\"uller-Hoissen,
 Differential calculi on commutative algebras,
 J. Phys. A: Math. Gen.
   28  (1995) 3197--3222.

\bibitem{DMH00CJP}
A.~Dimakis and F.~M\"uller-Hoissen,
 Pseudo-{R}iemannian metrics in models based on noncommutative geometry,
 Czech. J. Phys.
   50  (2000) 45--52.

\bibitem{Bach+Petr93}
C.~Bachas and P.M.S. Petropoulos,
 Topological models on the lattice and a remark on string theory cloning,
 Commun. Math. Phys.
   152  (1993) 191--202.

\bibitem{FHK94}
M.~Fukuma, S.~Hosono, and H.~Kawai,
 Lattice topological field theory in two dimensions,
 Commun. Math. Phys.
   161  (1994) 157--175.

\bibitem{DMH92_grav}
A.~Dimakis and F.~M\"uller-Hoissen,
 Noncommutative differential calculus, gauge theory and gravitation, 
 Report GOET-TP 33/92, http://wwwuser.gwdg.de/\~{
  }fmuelle/download/preprint92.ps (1992).

\bibitem{Dima+Tzan96}
A.~Dimakis and C.~Tzanakis,
 Non-commutative geometry and kinetic theory of open systems,
 J.Phys. A
   29  (1996) 577--594.

\bibitem{Kauf04}
L.H. Kauffmann,
 Non-commutative worlds,
 New J. Phys.
   6  (2004) 173.

\end{thebibliography}
\end{document}